\documentclass[10pt,twoside]{article}
\usepackage{amssymb}
\usepackage{latexsym}

\makeatletter

\@addtoreset{equation}{section} \makeatother
\newtheorem{theorem}{Theorem}[section]
\newtheorem{remark}[theorem]{Remark}
\newtheorem{lemma}[theorem]{Lemma}
\newtheorem{proposition}[theorem]{Proposition}
\newtheorem{corollary}[theorem]{Corollary}
\newtheorem{example}[theorem]{Example}
\newtheorem{definition}[theorem]{Definition}

\newcommand{\bdefi}{\begin{definition}}
\newcommand{\btheo}{\begin{theorem}}
\newcommand{\bprop}{\begin{proposition}}
\newcommand{\brema}{\begin{remark}}
\newcommand{\bcoro}{\begin{corollary}}
\newcommand{\blemm}{\begin{lemma}}
\newcommand{\bexam}{\begin{example}}

\newcommand{\edefi}{\end{definition}}
\newcommand{\etheo}{\end{theorem}}
\newcommand{\eprop}{\end{proposition}}
\newcommand{\erema}{\end{remark}}
\newcommand{\ecoro}{\end{corollary}}
\newcommand{\elemm}{\end{lemma}}
\newcommand{\eexam}{\end{example}}

\newcommand{\be}{\begin{equation}}
\newcommand{\ee}{\end{equation}}

\newcommand{\R}{{\mathbb R}}

\newcommand{\s}{\psi}
\newcommand{\cc}{c_{\bar \sigma}}

\newcommand{\ben}{\begin{enumerate}}
\newcommand{\een}{\end{enumerate}}
\newcommand{\bit}{\begin{itemize}}
\newcommand{\eit}{\end{itemize}}
\newcommand{\edoc}{\end{document}}

\title{
\vspace{0.5in} {\bf On the foundations and necessity of classical
gauge invariance}}

\author {\bf M. S\'anchez \\
{\it\small Departamento de Geometr\'{\i}a y Topolog\'{\i}a}\\
{\it\small Facultad de Ciencias, Universidad de Granada}\\
{\it\small Campus de Fuentenueva s/n, 18071 Granada, Spain} \\
{\small sanchezm@ugr.es }}

\begin{document}
\parindent=5mm
\date{}
\maketitle

\begin{quote}

\noindent {\small \bf Abstract.} {\small We argue that, ideally,
 the  ways  to measure magnitudes in   non-quantum theories of physics (spacetime, field theory),
limit drastically their possible mathematical models. In
particular, gauge invariance in the Yang-Mills framework, is a
necessity of our way of measuring rather than an {\em a priori}
imposition on symmetry.

A general postulational basis  for the geometric aspects of
classical field theories is introduced, and the permitted models
are studied.
 Some of them
(for example, compatible with signature-changing metrics or
variations of the speed of interactions) are new, and require a
generalization of the concept of principal fiber bundle, which may
be of interest both, physically and mathematically.
}\\

\end{quote}
\begin{quote}

\end{quote}
\begin{quote}

{\small\sl 2000 MSC:} {\small 70S15, 53C80.

}.
\end{quote}

\newpage
{\small \tableofcontents } 


\section{Introduction}



There is a long philosophical tradition which claims that, in
order to know the world, one has to study himself ---our own
structure reflects the world. In the accurate framework of the
theories of physical spacetimes,  Bernal, L\'opez and the author
\cite{BSfound} developed a precise variant of this claim:
 {\em our way of
measuring macroscopic space and time (summarized in three minimum
postulates), attaches one among four geometric structures to
physical spacetime}.

In the present article, our aim is to extend this viewpoint to
field theory showing, in particular, that  gauge invariance
emerges in classical field theories as a necessity of our way of
measuring
---not, say, as an a priori requirement of symmetry or as an
imposition of causal interactions. Certainly, this idea was
already suggested by the founders of the theory some decades ago
(see especially \cite{tH}). Nevertheless, we develop it in detail,
introducing a postulational basis, revisiting known arguments and
obtaining new possibilities.  So, our results can be summarized
as: (1) gauge invariance is a requirement of consistency for our
measures, not an ``optional'' mathematically elegant assumption,
(2) classical theories on spacetime, including not only General
Relativity but also Galilei-Newton one, can be regarded  as gauge
theories in the sense of Yang-Mills, and (3) there are some
natural possibilities beyond the standard framework of classical
gauge theories, which might be both, physically and mathematically
interesting.

Our approach is introduced in a more or less classic way. First,
the minimum consensus hypotheses on measures are postulated in a
mathematically rigorous way. Such postulates should be ``obviously
acceptable'' at least as effective or consensus claims. They are
even partially deducible from more elementary facts\footnote{That
is, they would be partially redundant, if one started at a more
elementary level. For example, as explained in Section \ref{s3},
the vector structure of the fiber $V$ in postulate (H1) can be
regarded just as a linear approximation, or the group action on
standard bases in (H2) appears necessarily because, otherwise, a
group action can be attached univocally.}. Nevertheless, the
``non-deducible'' part of these postulates sounds so elementary
that a universe where they did not held would seem radically
different to ours. In this sense, our conclusions can be useful
for present-day proposals such as the  Theories of Everything
(TOEs) or parallel universes (see for example Barrow \cite{Ba2}):
{\em our results bound the mathematical possibilities for
universes measurable in a way minimally similar to ours}.

Some  comments on this point are worth mentioning. To be more
specific, we will consider \cite{Te}, and retain part of his
terminology. In this recent reference (see also \cite{Te0}),
Tegmark has suggested a Mathematical Universe Hypotheses (MUH).
Accordingly, our Universe would be a mathematical
structure\footnote{Moreover, this structure would be one among
others, in a ladder of multiverses with  four levels of increasing
generality. Many physicists may feel reluctant or clearly opposed
to this type of ideas, and the public lecture by R. Penrose,
``Fashion, Faith, and Fantasy in Modern Physical Theories''
(delivered at 17th International Conference on General Relativity
and Gravitation) may be a prominent  example. Our approach is
classical and, thus, independent of such controversial topics
---our aim is not to dispute on them. We will regard ideas concerning parallel universes
as a natural background for compelling speculative items in
Physics, such as the many worlds interpretation of Quantum
Mechanics or inflation.},
free of our particular (cultural, biological) ``baggage''.
The necessity to descend from the (outside, mathematical) ``bird
view'' of the reality, to the (inside, experimentalist) ``frog
view''   is emphasized. To get this, Tegmark suggests to analyze
the symmetries of mathematical structures. In this paper, we
stress the reversed perspective, going from frog postulates
(expressed in a reasonably baggage-free way) to more general bird
views. Our conclusions yield hints for any TOE, as it must be
compatible with our conclusions (at least as a  non-quantum
limit). Moreover, they may suggest concrete ways to ``descend''
from the bird to the frog view, as suggested in \cite{BSS}.

This paper is organized as follows. In Section \ref{s2} we sketch
the approach for spacetimes in  \cite{BSfound}. As the study in
this reference is mathematically exhaustive, here we stress the
relevant physical  aspects. In the first subsection $\S$\ref{s2.1}
a brief account of our three postulates for spacetimes is given.
In the next one $\S$\ref{s22}, the four possible mathematical
structures derived from these postulates are explained. In the
last subsection $\S$\ref{s2.3}, we emphasize the existence of a
new element yielded by the theory. This is a function $k$ on the
spacetime which controls possible changes among the four
mathematical structures. Rigorously, if $k(p)$ is non-positive
then $c(p)=\sqrt{-k(p)}$ is the supremum of speeds between
standard observers at $p$. But $c(p)$ also admits a natural
interpretation as a (possible varying) speed of propagation of
interactions  --light-- in vacuum.

In Section \ref{s3}, three postulates for the geometric contents
of field theory are analyzed. The first and the third ones are
easy to understand, even though they are discussed in some detail
at the corresponding subsections: plainly, the first postulate
(H1) states that particle fields can be described by sections in
some fiber bundle space $E(M,V)$, $\S$\ref{s3.1}, and the third
(H3) is a technical claim on  the effective (macroscopic)
smoothness of the mathematical structures $\S$\ref{s3.3}. As in
the case of spacetimes, the second postulate (H2) is the crucial
one: it states the existence of standard bases for the observation
of (linear) fields at each event $p$. These bases are essential
for our way of measuring and, a posteriori, they come from the
automorphisms of the mathematical structures we are trying to
measure. The three postulates (H1), (H2), (H3) yield certain
fibred structure, obtained from the principal fiber bundle
$BE(M,V)$ of the bases of $E(M,V)$.  This is somewhat more general
than the usual structure in field theory. So, in $\S$\ref{s3.3},
we also discuss an additional postulate (H4)$^*$ which permits to
recover the familiar principle fiber bundle structure $P(M,G)$ of
standard Yang-Mills theories.

In Section \ref{s3.10} gauge invariance in the postulated
geometric framework for field theory is analyzed. Here we include
the additional postulate (H4)$^*$ in order to make the approach
directly applicable to the standard case ---but it will be removed
in Section \ref{s4.1}. In the first subsection $\S$\ref{ss2.3.1},
gauge invariance appears naturally from our way of measuring (not
as an ``a priori imposition of local symmetry'' from a global
one), and it is independent of the causal relations on the
spacetime. Careful mathematical distinctions among notions for
particle fields such as {\em gauge transformation, gauge orbit
{\em or} (gauge) naturally equivalent} particle fields,  are
introduced. These notions are equivalent in trivializable
principal bundles, but conceptually different. They become natural
when one considers a {\em standard field observer} as a section in
the fiber bundle $P(M,G)$ of standard bases on  $M$. Then, a {\em
principle of gauge invariance } emerges from this framework. In
the second subsection $\S$\ref{s3.2.2}, the known
 consequences of this principle are revisited. So, we discuss the necessity of a connection on $P(M,G)$ and
its possible interpretation as a particle field. In the last
subsection $\S$\ref{s3.2.3} we recall how all the spacetime
theories in Section \ref{s2} can be regarded also as field  and
gauge theories. This includes both, Galilei-Newton theory and
Einstein General Relativity. In principle, they appear as gauge
theories in the same footing, even though the interpretation of
the canonical Levi-Civita connection for the latter might differ
from the interpretation of the (non-unique) Galilean connection
for the former. We stress that, here, these spacetime theories are
regarded as Yang-Mills theories and, so, the structural group
appears on the corresponding fiber bundle for the spacetime (the
tangent bundle or associated tensor bundles); i.e., this group is
{\em not} the group of diffeomorphisms of the spacetime, as
frequently claimed for General Relativity (and critiqued in
\cite{We}).

In Section \ref{s4.1} the new possibilities which appear when
postulate (H4)$^*$ is not imposed, are explored. Roughly, this
means that the structural group $G$ of the classic principle
bundle $P(M,G)$ may vary with the event --for spacetimes, this is
 equivalent to assume that function $k(p)$ is not a constant, allowing variations in the speed of
 interactions. In the first subsection $\S$\ref{s4.1a}, we
explain the general mathematical structure $P(M,G_*)$ which
appears. Its technical difficulties are stressed with an example,
and are discussed in the second subsection $\S$\ref{s4.1b}. In the
last subsection $\S$\ref{s4.1c}, we check that the principle of
gauge invariance is extended naturally to this framework, and
leave its possible implications for future developments.

Some conclusions are summarized in the last section.

\section{Macroscopic space and time} \label{s2}

\subsection{The minimum consensus hypotheses on measures}\label{s2.1}

Historically, the most relevant physical theories of spacetime are
Galilei-Newton Classical Mechanics (with or without ``external
ether'') Special and General Relativity. As emphasized in
\cite{BSfound}, all of them share some minimum consensus
hypotheses on how space and time must be measured. These
hypotheses are valid not only  for these theories but also, in
principle, for  conceivable measures of a continuum macroscopic
spacetime (including even gedanken measures). Concretely, the
postulated hypotheses are:

\begin{enumerate}

\item[(P1)] The existence of a set $M$ of {\em events} (``here and
now'') which can be suitably labelled by using four coordinates
$(t, x^i), i=1,2,3$, the first one (``time coordinate'') clearly
distinguishable from the other three (spatial coordinates). In
particular, $M$ is endowed with a structure of smooth connected
manifold.

\item[(P2)] The possibility to find, at least infinitesimally,
{\em standard observers} (or, more properly,  {\em standard
observations} realized as coordinate charts) around each event
$p\in M$. These are characterized by the following minimum
symmetry assumption:

Any two charts $O\equiv (t, x^i)$, $\tilde O\equiv (\tilde t,
\tilde x^i)$ obtained by standard observers around $p$ satisfy
\begin{equation} \label{ep2}
\partial_{\,t}\,{\tilde t}|_p=\partial_{\,{\tilde t}}\,t|_p \; , \,\;\,\,\;\,
 \partial_{\,x^j}{\tilde x}^i|_p=\partial_{\,{\tilde x}^i}x^j|_p \; ,\,\;\forall i,j\in\{1,2,3\}.
\end{equation}

\item[(P3)] The effective smoothability of the possible geometric
structures assigned to $M$ by means of the previous item (P2)
---as well as by means of item (P1), but recall that smoothness
was already claimed explicitly there.

\een

These minimum consensus hypotheses are widely discussed in
\cite{BSfound}, where they are introduced rigourously. The first
and last one are very easy to understand (and accept). For the key
second one (P2), just recall:

\bit \item Implicitly, (P2) is assumed when one speaks on ``freely
falling observers'' in General Relativity, or on ``inertial
observers'' in Special Relativity and Classical Mechanics. In the
latter case (P2) would hold even if a sort of ether (which selects
more restrictively which observers are standard) were assumed.

\item Equalities (\ref{ep2}) constitute a minimum symmetry
assumption which means that, given two such standard observers
$O$, $\tilde O$,
 the comparison between their temporal (resp. spatial) coordinates at $p$ cannot privilege any of them.

 In fact, $\partial_{\,t}\,{\tilde t}|_p=\partial_{\,{\tilde
t}}\,t|_p$ means that the ${\tilde O}$-time, measured with the $O$
clock, goes by as the $O$-time, measured with the ${\tilde O}$
clock (this is just a sensible mathematical translation of the
assertion: ``$O$ and ${\tilde O}$ measure at $p$ using the same
unit of time''). For example, in Classical Mechanics there is a
sort of ``absolute time'', and standard observers measure them
obtaining $
\partial_{\,t}\,{\tilde t}|_p= 1 = \partial_{\,{\tilde t}}\,t|_p.$
Otherwise, the  time observed by  $\tilde O$ will present some
``time dilation'' (or contraction) $\partial_{\,t}\,{\tilde t}|_p$
with respect to  $O$; then, we postulate that $O$-time will
present an equal time dilation  for $\tilde O$.

In a similar way,  $ \partial_{\,x^j}{\tilde
x}^i|_p=\partial_{\,{\tilde x}^i}x^j|_p$ means that  the $i$--th
spatial unit of ${\tilde O}$, measured with the $j$--th ruler of
$O$, is identical to the $j$--th spatial unit of $O$, observed
with the $i$--th ruler of $\tilde O$. Again,  this is a sensible
mathematical translation of the assertion: ``$O$ and ${\tilde O}$
measure at $p$ using the same  units of space''.
 \eit
Summing up:
\begin{quote}
i) The second  postulate (P2)  seems to be an unavoidable symmetry
in our {\em frog} perception of spacetime, where time and space
are clearly different (and, consequently, the symmetry for time
and space measures are a priori independent). This symmetry is
expressed in a purely mathematical (``baggage free'') way and,
thus, seems appropriate for searching a more general theory.

ii) Postulates (P1), (P2), (P3) have been minimum consensus
hypotheses on how space and time are measured, as reflected for
the fact that, historically, all the physical theories on
continuum spacetime include them. But, moreover, they are so
fundamental that seem unavoidable in any macroscopic theory of
spacetime similar to ours.
\end{quote}

\subsection{The set of possible mathematical models} \label{s22}

As proved in detail in \cite{BSfound}, the careful analysis of the
three postulates above, show the possibility to assign one of some
(few) mathematical structures to the set of events $M$. For
completeness, we sketch in the Appendix how to do this, and we
only describe the final results next. Let $S^1$ be the
circumference obtained from the extended real line $[-\infty,
\infty]$ by collapsing $\pm \infty$ to a single point $\omega$,
and let $O^k(4,\R)$ be the groups defined around (\ref{ok}),
(\ref{okprima}) in the Appendix. Notice that, when $k\in (-\infty,
0)$, then  $O^k(4,\R)$ is conjugate to Lorentz group (in fact, it
becomes the Lorentz group for Lorentz transformations with speed
of light $c=\sqrt{-k}$), and  when $k\in (0,\infty,)$ then
$O^k(4,\R)$ is conjugate to the orthonormal Euclidean group. If
$k=\omega$ then $O^k(4,\R)$ is the Galilei group, and if $k=0$ it
is a mathematically dual group. Now, it is possible to prove:

\bit \item[(A)] For each event $p\in M$, some $k(p)\in S^1$ can be
assigned. Such a $k(p)$ either (i) it is univocally determined at
$p$, and the group $O^{k(p)}(4,\R)$ is assigned to $p$,
or\footnote{\label{f2} There exist a third residual possibility,
namely, the existence of (at most) four values of $k(p)$ all of
them positive. Even though this possibility can be handled without
any problem \cite[Subsect. 4.3 ]{BSfound}, it disappears under
very mild additional hypothesis \cite[Sect. 5 ]{BSfound} and is
scarcely representative (its appearance is due to the existence of
``only a few'' standard observers, with a special
 symmetry among them). So, we will not take into account this.} (ii) it can be chosen arbitrarily at $p$, and
$\{1\}\times O(3,\R) \equiv \cap_{k\in S^1} O^k(4,\R)$ is assigned
to $p$. The way to assign such a $k(p)$ and the corresponding
group is the following. Take the set of standard observers at $p$,
and the corresponding set of bases  $B_p= (\partial_t|_p,
\partial_i|_p)$ of the tangent space $T_pM$ induced by these standard observers.
Then, the transition matrix between two such bases belong to the
group $O^{k(p)}(4,\R)$ (see the Appendix).

\item[(B)] Now, apart from the function $k(p)$ on all $M$, we can
assign to $M$: \ben \item In the points where $k(p)\in (-\infty,
0)$: a Lorentzian metric, as in General Relativity.  \item In the
points where $k(p)= \omega$: a ``Leibnizian structure'', which is
a big generalization of classical Newtonian structures.
Concretely, such a structure consists of a  non-vanishing one form
$\Omega$ and a (positive definite) Riemannian metric $h$ on its
kernel (see \cite{BS} for an exhaustive study). In the classical
case, there exists a ``time function'' $t$ and $\Omega=dt$.
(Notice that, as a difference with the semi-Riemanian case, such a
structure does not select any connection; nevertheless, a
connection must be selected under a gauge principle as below, see
Subsection \ref{s3.2.3}.)

\item In the points where $k(p)= 0$: an ``anti-Leibnizian
structure'', which is a dual version of Leibnizian
ones\footnote{Concretely, an anti-Leibnizian structure is a
non-vanishing vector field $Z$ and a Riemannian metric $h^*$ on
its kernel in dual space.}.

\item In the points where $k(p)\in (0, \infty)$: a Riemannian
metric.

\item In the points where $k(p)$ takes all the values of $S^1$:
any of the previous structures (as well as others). Nevertheless,
this case cannot hold on some parts of $M$ at the same time that
any of the previous ones, if postulate (P3) is applied in a strict
sense. In fact,  under a smooth variation of the group assigned by
(A) at each point (i.e., either one of the 6-dimensional groups
$O^{k(p)}(4,\R)$ or the 3-dimensional $O(3,\R)$), the dimension
should vary continuously, that is, it must be constant   (see also
Section \ref{s4.1}). Therefore, the case $O(3,\R)$ will be
dropped, taking into account that, on one hand, we have strong
experimental evidences that $k(p)$ must be no-negative in some
parts of our Universe, and, on the other, the case $O(3,\R)$ can
be regarded as a special limit of the generic case
$O^{k(p)}(4,\R)$. \een \eit Summing up:
\begin{quote}
The minimum postulates (P1), (P2), (P3) about how space and time
are measured (in a ``macroscopic way'') imply that we can assign
to spacetime either one of the four mathematical structures 1--4
above  (or, eventually, the ``degenerate'' fifth one), or a
structure on $M$ varying smoothly with the point among these four
ones.
\end{quote}
That is,  only these four structures are the result of the
effective symmetries perceived from our frog viewpoint. If a TOE
existed or Tegmark's MUH were true, such effective symmetries
should be explained as partial symmetries or limit symmetries
derived from the bird viewpoint. And, at any case, they must be
taken into account in any description (even if less ambitious than
a TOE) of our space-time reality.

\subsection{The role of $k(p)$ and the speed of light}\label{s2.3}
Some words on the role of function $k(p)$ are in order. Recall
first:

\ben \item[(i)] When $k(p) \not\in (0,\infty)$ then
$c(p)=\sqrt{-k(p)}$ admits the natural interpretation of {\em
supremum of relative velocities between standard observers}.

The case $k(p) \in (0,\infty)$ cannot hold if we assume the {\em
Postulate of Temporal Orientation } (PTO), that is,
$\partial_{\,t}\,{\tilde t}|_p
> 0$ hold for standard observers around\footnote{\label{f6}Technically, this must hold not only for the set $S_p$
of standard observers around $p$ but also for the set $S^*_p$ of
observers which share the symmetries (\ref{ep2}) with all the
observers of $S_p$ and, thus, can be also regarded as standard,
see \cite[Defn. 2.1]{BSfound}.} $p$. This postulate would be
clearly natural in the Universe around us.

\item[(ii)] If PTO were accepted, other assumptions might be
reasonable. A, say, {\em Postulate of Electromagnetic
Interpretation} (PEI)\footnote{The justification of this postulate
(under the previously accepted (P1), (P2), (P3), (PTO)), in a
reasonably general  way (with ``no heavy baggage''), would come
from the experimental input: (1) light propagates at a finite
speed  in vacuum, and (2) vacuum looks like equal for all standard
observers around each event $p$. As a consequence, the measured
speed of light $c_p$ at each $p$ must be equal for all the
standard observers. As the unique scalar quantity at each $p$ from
previous postulates is $c(p)$, this leads to identify $c(p)=c_p$.
}
would state that $c(p)$ is equal to the speed of light at $p$ and
$0<c(p)<\infty$. If this were accepted then one can try to justify
also the {\em constancy of the speed of light} $c(p)\equiv c\in
 (0,\infty)$ (independent of $p$) as a general fact.
\een Thus, one is tempted to include a fourth  postulate for
spacetime which, under PEI, would express the constancy of light
speed:

\begin{enumerate}
\item[(P4)$^*$] Function $k(p)$ is a negative constant $k=-c^2$
independent of $p$ (and, thus, a unique  group $O^{k}(4,\R)$ must
be considered on all $M$). \een

Nevertheless, some caution with (P4)$^*$ or the other ``additional
postulates'' is needed, when one is looking for a general (bird)
theory. All these postulates (as well as (P1), (P2), (P3) above)
collect perceptions from our frog perspective: if a TOE (or
reasonably general theory) existed, it should be compatible with
them. But we are not aware on how strange a TOE may seem. So, our
frog postulates must be consensus hypotheses as minimal as
possible. They must  hold in the (relatively small) part of the
Universe we can measure, but also allow generality for possible
extrapolations. In this sense, it is natural to think that the
inequality stated by PTO will hold in vast regions of our
Universe, but not necessarily on all it. A general
signature-changing metric (from Lorentzian to Riemannian, with
degenerate parts which eventually may be Leibnizian or
anti-Leibnizian), in the spirit of the limit of Hartle and Hawking
proposal \cite{HH} (see also, for example, \cite{Dr, MSV} and
references therein), is fully compatible with our basic three
postulates and our limited experimental observations. Moreover,
the  evidences of, say, the constancy of the speed of light, might
be non so clear (as  claimed sometimes) and affect several
measures ---for example, the accelerated expansion of the
Universe.
This question might be a testable (experimentalist,
frog) problem.

Summing up:
\begin{quote}
If (P4)$^*$ is assumed (in addition to (P1), (P2), (P3)) then the
mathematical ambient for our description of physical spacetime is
a Lorentzian 4-manifold, as in classical General Relativity.

If (P4)$^*$ is not assumed, other possibilities, as the existence
of a signature changing metric or the variation of the speed of
light, appear (Subsection \ref{s4.1}).
\end{quote}

\section{Field   theories} \label{s3}

Along this section  we will define a general framework for field
theory, based again in hypotheses on our way of measuring (close
to the observer) and as ``baggage free'' as possible. As in the
case of spacetimes, we will introduce three basic ``minimum
consensus hypotheses'', or postulates. Finally, we discuss a
fourth one, similar to (P4)$^*$ for spacetimes, which   simplifies
the mathematical approach and recovers the accepted framework for
classical Yang-Mills theories.

As in the case of spacetimes, we focus only in the geometric
aspects on measures. That is, we will not consider essential
ingredients such as energy, Lagrangians or equations of evolution.

\subsection{Ambient Hypotheses (H1)} \label{s3.1}
Let us start finding  some consensus hypotheses on our way of
measuring in field theory. We will consider as a first set of
hypotheses the following postulate:

\ben \item[(H1)] (Framework) (a) The mathematical ambient for
field theory is a fiber bundle $E(M,V)$, where a set of primary
physical fields on $M$ is represented by a section of the bundle
$\s:M\rightarrow E$, or {\em particle field}.

(b1). For the base $M$ (the {\em underlying spacetime}), the
postulates (P1),(P2), (P3) in Section \ref{s2} hold.

(b2). The fiber $V$ (the {\em model target space} for the values
of the physical magnitudes at each event) is a 
vector space of finite dimension $m$,  
and the fiber bundle $E(M,V)$ is a (real) vector bundle. \een

Notice that the part (a) only provides the most general ambient in
Differential Geometry; its role is similar to (P1) in Section
\ref{s2}. The word ``primary'' is introduced to keep track of the
fact that, by developing the theory,  new physical fields may
appear. 
Notice also
 that  two different sections $\s, \bar \s$ may describe the
same physical fields, as discussed in Section \ref{s3.10}.

 The assertion about
 $M$ in (b1) only states the compatibility with our study of spacetime in
previous section. The stated properties on the fiber $V$ in (b2)
are a simplification, but there are  reasons to assume it, at
least for
 a ``effective but general'' 
 theory:

(i)  Finite dimensionality would be natural for our capability to
measure only a finite number of variables, and

(ii) The linear character of the fiber $V$ is justified by the key
role of  linear mathematical approximations. That is, perhaps a
more subtle fiber fitted better, but we can measure directly only
linear approximations.

\smallskip
\noindent Moreover if, say, $V$ were a complex  vector space, it
could be also regarded as  a real one, and this is a natural
choice for the vector bundle structure, as the base is real.

There are  two more possibilities of interest for the fiber, even
accepting the arguments (i) and (ii) above: (a) to assume that $V$
has more than one connected component (say, in order to describe a
discrete variable), or (b) to assume that $V$ is an affine space,
with the 0 section not defined a priori (a candidate for such a
section might appear a posteriori after a sort of ``spontaneous
symmetry breaking''). But even these cases can be handled in a
similar way. 
Summing up, we maintain in what follows for simplicity (or, at
least, as a first approximation), that $V$ is a (finite
dimensional) real vector space.

\subsection{Standard bases (H2)}\label{s3.2}

For any vector bundle $E(M,V)$, 
one can construct the manifold $BE$ consisting of all the bases of
the fiber $E_p$ at any point $p$ of $M$. This turns out a
principle fiber bundle $BE(M,G_m)$, where the structural group
$G_m$ is just the general linear group $G_m=$ Gl$(m,\R)$. $G_m$
acts on the right naturally on each fiber $BE_p$: if $u=(e_1,\dots
, e_m) \in BE_p$ and $g\in G_m$ then $ug$ is obtained just by
multiplying formally. The following hypothesis plays a role
similar to (P2) for spacetimes.

\ben \item[(H2)] (Standard bases). For each $p\in M$, our way of
measuring the physical fields around $p$ selects a proper subset
$P_p$ of $BE_p$ ($P_p\neq \emptyset, BE_p$), whose elements will
be called {\em standard bases at $p$} and satisfy:

There exists a closed subgroup $G_p \subset G_m$ which acts
freely and transitively on $P_p$.
 \een
Notice that, chosen a bases $u$ in $P_p$, the claimed action
allows to identify $P_p$ and $G_p$,
but this identification depends on $u$.

Recall that (H2) has been historically a minimum consensus
hypothesis on our way of measuring, including the standard model
of particle physics. For example, the bases in $P_p$ may be
orthonormal for a real or hermitian product, and $G_p$ is then the
set of orthonormal or unitary matrixes. But the {\em a priori}
reasonability of (H2) comes from the facts: \ben \item[(i)]  $P_p$
must be a proper subset of $BE_p$: if no standard bases existed at
$p$ (or if all the bases were standard) there would not be  ways
to specify any intrinsic property of a vector $v$ in $E_p$, except
if $v$ is 0 or not.

\item[(ii)] A group $G_p$ acts on each fiber $E_p$: this is
commonly assumed as a natural requirement of symmetry (even more,
with $G_p$ independent of $p$), but a more precise justification
is the following. Assume first that $P_p$ is any (non-empty) set
of bases of $E_p$. Choose $u\in P_p$ and define $G_{p,u}=\{g\in
G_m: ug\in P_p\}$. The smallest subgroup $G_p^*$ of $G_m$ which
contains $G_{p,u}$ is independent of $u$. This unique subgroup
$G_p^*$ is determined from $P_p$ and acts freely and transitively
on some set of bases $P_p^* \supset P_p$ (also univocally
determined). So, one can assume that the extended set of bases
$P_p^*$ will play the role of standard bases. \een
\noindent As a more abstract argument, any intrinsic property for
vectors in the fiber $E_p$ can be described by taking the group of
automorphisms for this property. This group will act on $E_p$, and
a class of bases related by this action will be the set of
standard bases.

It is worth comparing (H2) and (P2):

\bit \item The assumption (H2) on the action of a group between
standard bases only expresses that there will be some symmetry
between these bases. In (P2) we did not assume the action of any
group, but we deduced its existence\footnote{Some technical
details may be taken into account. This action is obtained not for
the (rather arbitrary) original set $S_p$ of standard observers
but for a natural set $S^*_p$ constructed from the original one
(see footnote \ref{f6}), which plays a similar role of $P_p^*$
above. Nevertheless, the result for spacetimes is sharper, as one
proves the possibility to relate the bases of $S_p$  with elements
of some of the groups $O^k(4,\R)$. In all the cases, but in the
residual one (footnote \ref{f2}) $S^*_p$ appears naturally as
$P^*_p$ above (in fact, either $k$ is univocally determined, and
$O^k(4,\R)$ acts on $S^*_p$ or $k$ can be chosen  arbitrarily, and
$O(3,\R)$ acts). In the residual case, one can also define $S^*_p$
from a minimum group $G_p$ (as done above with $P^*_p$). But there
exists also the possibility to determine univocally (at most) four
values of $k$. The corresponding groups $O^k(4,\R)$ would be
subgroups of the minimum group $G_p$. However, this case disappear
under hypotheses very scarcely restrictive \cite[Section
5]{BSfound}), and the present discussion is only an academical
one.} and, then, (H2) will also hold in this case. \item (H2)
states that some symmetries among standard bases will exist, but
we do not assume a priori anything about this symmetry. On the
contrary, in (P2) we used the concrete symmetry (\ref{ep2}),
motivated by our familiar distinction between time and space. \eit

\subsection{Effective smoothness (H3), constancy of the structural group
(H4)$^*$}\label{s3.3}

\noindent Now, we can state that smoothness is an effective
macroscopic approximation, as claimed in  (P3):

\ben  \item[(H3)] (Smoothability) All the  geometric structures
obtained  by means of the previous item (H2) (as well as for the
item (H1)) are (differentiably) smooth. In particular,
$P=\cup_{p\in M} P_p$ is a smooth submanifold of $BE(M,G_m)$.

\een Notice that this differentiability allows $G_p$ vary from one
point to another, even though this variation must be smooth. We
will discuss discuss precisely what differentiability implies in
Section \ref{s4.1}; in particular, the dimension of $G_p$ will be
regarded constant.

 The consensus hypotheses (H1), (H2), (H3) are natural extensions of (P1), (P2) and
 (P3).
And  we can wonder, extending in Subsection \ref{s2.3},  at what
extent the assumption that $G_p$ is independent of $p$ is
reasonable. The constancy of $G_p$ with $p$ is not as compelling
as the constancy of its dimension stated above neither, in
general,  as (H1), (H2), (H3). With this caution, we introduce:

\begin{enumerate} \item[(H4)$^*$]  There exists a closed subgroup $G\subset G_m$ such that $G_p=G$ for all $p\in M$. \een

We emphasize that there are as many  evidences for (H4)$^*$ as for
(P4)$^*$ above, that is: there is no experimental evidence against
them. However, the consensus of the physical community for
(H4)$^*$ might be bigger. The reason is that,  in the case of the
spacetime, there exist theories which admit the signature change.
Moreover, under our approach the variation of $c(p)$ would admit
(in principle)  testable evidences. But, as far as the author
knows, there is nothing analogous in the case of field theory.
Nevertheless, the variation of $G_p$  may be a possibility worth
of exploring theoretically, which will be discussed again in
Section \ref{s4.1}.

Recall that (H4)$^*$, in addition to previous hypotheses, yields a
structure of principal fiber bundle on $P$. Thus, summing up, we
have obtained:

\begin{quote}
If (H4)$^*$ is assumed (in addition to (H1), (H2), (H3)) then the
mathematical ambient for our description of field theory is a
principal fiber bundle $P(M,G)$, obtained as a reduction
(subbundle) of the bundle of the bases $BE(M,G_m)$; its base $M$
represents the underlying spacetime, and the fiber $G$ is a
(finite dimensional) Lie group.

If (H4)$^*$ were not assumed, another possibilities of fiber
bundles would appear (Subsection \ref{s4.1}).
\end{quote}
Two technical notes. First, notice that $E(M,V)$ can be recovered
from $P(M,G)$ as an associate vector bundle (see for example
\cite[p. 30]{Po}). Second,  once the structure of principle fiber
bundle for $P(M,G)$ is accepted, one can construct an associate
vector fiber bundle $E^\rho(M,V)$ for any representation $\rho:
G\rightarrow $Aut$V$, as commonly used in particle physics for
faithful representations. Even though technically useful, in
principle, this does not yield more generality  from the
fundamental viewpoint.

\section{Gauge Theory}
\label{s3.10}

Since the seminal paper by Yang and Mills \cite{YM}, gauge
principles and their interpretations in terms of connections in
fiber bundles are very well-known (see for example, \cite{Bl},
\cite{Na}). Now, we will revisit the fundamental physical ideas.
We will show that our minimum hypotheses in last section lead to a
principle of gauge invariance. Thus, this principle can be also
regarded as a minimum
consensus hypotheses.

Next, we will assume that postulates (H1), (H2), (H3) hold. For
simplicity, (H4)$^*$ will be also assumed here --as always in
classical Gauge Theory--, and the possibility to remove it will be
explored in the next section. So, next $P\equiv P(M,G) \subset
BE(M,G_m)$ is a principal bundle, as explained above.

\subsection{Gauge invariance} \label{ss2.3.1} Our aim is to show that, given
any particle field $\s: M\rightarrow E$, there exists a set of
particle fields, the gauge orbit of $\s$, Orbit$(\s)$, which
cannot be distinguished of $\s$ by any experimental method. This
justifies the principle of gauge invariance, which will assert
that all these particle fields describe the same physical reality
and, thus, will yield the same physical quantities.

We start by working in a {\em trivializing neighbourhood}
$U\subset M$ for $P$, that is, $U$ satisfies that $P(M,G)$ admits
a section on $U$ and, thus, the restrictions $P^U\equiv P(U,G)$
and $E^U\equiv E(U,V)$ are trivializable bundles.

Ideally, in order to measure a particle field, an observer should
take a standard basis at each event $p$ defining some section
$\sigma: U\subseteq M \rightarrow P$ of the principal bundle
$P(M,G)$. Then, at each $p\in U$, the basis $\sigma(p)$ will yield
some coordinates $c(v_p)\in \R^m$ for any $v_p \in E_p$. Regarding
$\sigma(p)$ as a $m$-uple of vectors, $\sigma(p)= (e_1(p), \dots ,
e_m(p))$, and $c(v_p)$ as a column vector of $\R^m$, one obtains
naturally a smooth map of coordinates on $E^U$ characterized by:
\be \label{ecoor} c: E^U \rightarrow \R^m, \quad \quad v_p =
\sigma(p) c(v_p). \ee

\bdefi (1) A {\em standard (field) observer} is any section
$\sigma: U \rightarrow P$.

(2) Its set of {\em associate coordinates} $c (\equiv c_\sigma)$
is the function defined by (\ref{ecoor}).

(3) The {\em coordinates of a particle field $\s: M \rightarrow E$
measured by the standard observer $\sigma$} is the composite
function $c \circ \s: U\rightarrow \R^m$.
 \edefi
Of course,  the coordinates of a particle field change with the
standard observer. If $g_U: U\rightarrow G$ is any smooth map then
the section $\bar\sigma: U\subseteq M \rightarrow P, p
\rightarrowtail \sigma(p)\cdot g_U(p)$ also defines a standard
observer with coordinates
$$
\cc = g_U^{-1} c
$$
(notice that each $g_U(p)$ belong to a group of matrixes); one may
say that the new coordinates $\cc \circ \s: U\rightarrow \R^m$ for
the particle field $\s$ are obtained from coordinates $c \circ \s$
by means of a ``passive pointwise symmetry''.

Two  different standard observers $\sigma, \bar\sigma $ may assign
the same coordinate function to two particle fields $\s, \bar \s$
(the particle field $\bar \s$ is obtained from $\s$ by means of an
``active pointwise symmetry''). In this case, the fact that {\em
standard observers $\sigma, \bar \sigma$ must be physically
equivalent should imply that the two particle fields are
physically indistinguishable on $U$}. This will be claimed below
as a physical property, but let us introduce it progressively.

\bdefi \label{dident} Let $\s, \bar \s: M\rightarrow E$ be two
particle
 fields which satisfy:  for each standard observer $\sigma: U \rightarrow P$ there exists
 another standard observer $\bar \sigma: U \rightarrow P$ such
 that the corresponding coordinate functions of the particle fields coincide, i.e.,
 $c_\sigma \circ \s = c_{\bar \sigma} \circ \bar \s$.

Then, $\s$ and $\bar \s$ are called {\em naturally
indistinguishable.} \edefi

 \brema \label{rpgi} {\em Trivially  the binary relation
 ``to be naturally indistinguishable'' in the set of all the particle
 fields is transitive (and  a relation of equivalence).
 }\erema
On a trivializing open subset $U$ for $P$, we can  characterize
easily all the particle fields which are naturally
indistinguishable to a prescribed one $\s_U:U\rightarrow E$ as
follows. Fix a standard observer $\sigma_0: U\rightarrow P$.
Associated to $\sigma_0$ one has an action $\cdot$ of $G$ on $E^U$
defined as: \be \label{eact} g \cdot v_p = \sigma_0(p) g c_0(v_p)
, \quad \quad \forall g \in G, \forall v_p\in E^U, \ee where $c_0$
is the set of associated coordinates for $\sigma_0$.

For any fixed $g\in G$, the new section $g^{-1} \cdot \s_U$, which
is naturally indistinguishable to $\s_U$, is called classically a
{\em global gauge transformation on $U$ of $\s_U$}. Moreover, for
any smooth map $g_U: U\rightarrow G$, the {\em
(pointwise)\footnote{The usual name is ``local'' gauge
transformation, but we will not use it, in order to avoid
confusions with local properties such as the trivialization of
fiber bundles.} gauge transformation}

\be \label{egg} \bar \s_U(p)= g_U^{-1}(p) \cdot \s_U(p) \quad
\quad \forall p\in U \ee is also naturally indistinguishable to
$\s_U$. The {\em gauge orbit on $U$ of } $\s_U$, Orb$_U(\s_U)$ is
the set of all such gauge transformed fields, obtained for any
$g_U: U\rightarrow G$. By construction, Orb$_U(\s_U)$ is the set
of all the particle fields on $U$ naturally indistinguishable to
$\s_U$.

Now, let us extend gauge transformations on trivializing $U$ to
all $M$. Consider a particle field defined as a global section
$\s: M \rightarrow E$. For each trivializing open set $U\subset M$
of $P(M,G)$, choose a section and the associated action
(\ref{eact}). Consider functions $g_U: U\rightarrow G$ such that
$g_U$ is identically equal to the identity matrix $I_m\in G$ on a
neighborhood of the boundary $\partial U$ of $U$ in $M$. Clearly,
the particle field: \be \label{elg} \bar\s (p) = \left\{
\begin{array}{ll} g_U^{-1}(p) \cdot \s(p) &
\forall p \in U \\
\s (p) & \forall p \in M\backslash U.
\end{array}
\right. \ee is well defined on all $M$.
\bdefi \label{dorbit} (1)
A {\em gauge transformation} of the particle field $\s: M
\rightarrow E$ is any particle field $\bar \s$ obtained from
(\ref{elg}) for some trivializing $U$ and some function $g_U$
equal to the identity in a neighborhood of $\partial U$.

(2) The {\em gauge orbit} Orb$(\s)$ of $\s$ is the smallest set of
particle fields which satisfies:

(i) $\s \in$ Orb$(\s)$ and,

(ii) if $\bar \s \in$ Orb$(\s)$ then any gauge transformation of
$\bar \s$ belongs to Orb$(\s)$.
 \edefi
\brema \label{rorbit} {\em (1) If the principle fiber  bundle
$P(M,G)$ is trivializable, then Orb$(\s)$ is just the set of all
the gauge transformations of $\s$. Otherwise, they are not equal.
In fact, any such gauge transformed $\bar \s$ coincides with $\s$
at some points: as no global section exists, there exists some
$p_0\in M$ (any point not included in the domain of the section)
and a neighborhood $U_0\ni p_0$ such that $\s(p) =\bar \s(p)$ for
all $p\in U_0$. So, in the non-trivializable case, the binary
relation ``$\bar \s$ is related by means of a gauge transformation
with $\s$'' on the set of all particle fields, is not transitive
(compare with Remark \ref{rpgi}). Nevertheless, to belong to the
same orbit is obviously a relation of equivalence. Moreover,
Orb$(\s)$ can be constructed as follows. Consider the smallest
relation of equivalence  which contains the binary relation ``to
be gauge related'' (i.e., the intersection of all the relations of
equivalence which contain the binary relation induced by Defn.
\ref{dorbit}(1)). Then, Orb$(\s)$ is the class of equivalence of
$\s$.

(2) Anyway, the smallest set which defines Orb$(\s)$ can be
constructed explicitly as the union $\cup_{n=0}^\infty $
Orb$_n(\s)$, where each Orb$_n(\s)$ is defined recursively as
follows:

(i) Orb$_0(\s)=\{\s\}$,

(ii) Orb$_n(\s)$ is the set of all the particle fields obtained as
a gauge transformation of some particle field in Orb$_{n-1}(\s)$.
\smallskip

\noindent Thus, if $\s, \bar \s $ belong to the same orbit, there
is a finite chain of gauge transformations which sends the first
field  into the second one.}\erema


 \btheo If two particle fields $\s,
\bar\s: M \rightarrow E$ belong to the same gauge orbit
(Orbit($\s$)=Orbit($\bar\s$)) then they are naturally
indistinguishable. \etheo

\noindent {\em Proof.} From Remark \ref{rorbit}(2) there exists a
chain of particle fields $\s_0= \s, \s_1, \dots, \s_k=\bar \s$,
such that each two consecutive fields are related by a gauge
transformation and, thus, are naturally indistinguishable. As this
relation is transitive (Remark \ref{rpgi}) the result follows.
$\Box$


\brema {\em Along this section, we have considered three
conditions on particle fields $\s, \bar \psi$:

(a) To be related by the a gauge transformation (Defn.
\ref{dorbit}(1)).

(b) To lie in the same gauge orbit  (Defn. \ref{dorbit}(2)).

(c) To be naturally indistinguishable (Defn. \ref{dident}).

\smallskip

\noindent If $P(M,G)$ is trivializable (as happens necessarily if,
for example, $M$ is contractible), these three conditions
coincide. But in general one only has (a) $\Rightarrow$ (b)
$\Rightarrow$ (c).

In principle, this distinction might be regarded as a mathematical
subtlety which does not affect the essence of our physical
discussion. Nevertheless, we will take it into account below not
only in order to be totally accurate but also to bear in mind that
we are dealing with  three different concepts ---apart from the
fact that non-local experimental effects type Ahanorov-Bohm may
stress its physical importance. }\erema Up to now, a name such as
``naturally indistinguishable'' has been introduced just as a
mathematical definition, being the name only suggested by other
more elementary ones such as ``standard observer''. But in order
to make a physical theory we must postulate in a precise way at
what extent this mathematical definition corresponds with reality.
Taking into account the discussion above Defn. \ref{dident},  if
$\s$ and $\bar \s$ are naturally indistinguishable then no direct
measure by standard observers can distinguish between them.
Nevertheless, (in the non-trivializable case) no domain $U$ of any
standard observer covers all of $M$. Thus, in principle, the
possibility that an indirect measure of some global property on
$M$ distinguished them, must be taken into account. Of course,
this would not be the case if $\s$ and $\bar\s$ are related by a
gauge transformation. In this case, the possible differences
between the two fields would be measurable in the region where $\s
\neq \bar\s$. But the equivalence between standard observers must
imply that both particle fields represent the same reality. That
is, the physical fields described by them must generate the same
measurable magnitudes and cannot be distinguished by any
experimental method. Nevertheless, this is not sufficient yet. It
is obvious that the relation ``to represent the same physical
fields'' is a relation of equivalence, but to be gauge related is
not, recall Remark \ref{rorbit}. Thus, particle fields in the same
gauge orbit must also represent the same physical fields. So, we
arrive naturally to the following consensus hypotheses about our
measures:

 \ben \item[(PGI)$^*$] (Principle of Gauge Invariance). Under postulates (H1), (H2), (H3), (H4)$^*$,
all the particle fields in the same gauge orbit Orb$(\s)$ are {\em
physically identical}, that is, all of them describe the same set
of (primary) physical fields.
 \een
\begin{remark}{\em  Notice that we have ``deduced'' (PGI)$^*$ from simple interpretations
about standard field observers. In order to avoid a formalization
of these interpretations, we state (PGI)$^*$ as a new postulate.
But we emphasize
that {\em (PGI)$^*$ emerges as a requirement of our way of
measuring, not as an a priori assumption on symmetry.}
}\end{remark}

\subsection{Necessity of fiber connections and gauge fields}
\label{s3.2.2} Now, we revisit the classical implications of
(PGI)$^*$ on the necessity of gauge fields. Under the classical
viewpoint on field theory, one assumes the existence of a
Lagrangian density ${\cal L}$ defined on the space of 1-jets
$J^1(E)$ (roughly, each element of this space gives a point $p$ of
$M$, an element $v_p$ of the fiber $E_p$, and the differential
$\theta$ of some section which sends $p$ to $v_p$). Given a
particle field $\s: M\rightarrow E$ which describes the primary
physical fields, ${\cal L}$ can be applied to the the section
$d\s$ of 1-jets induced from $\s$, defining so a function ${\cal
L}(d\s): M\rightarrow \R$.

Fix a standard observer $\sigma_0: U\rightarrow P$ and the
associate action (\ref{eact}) on $E^U$. This action  also induces
a natural action  on $J^1(E^U)$. As $G$ is related to symmetries
of the physical system, the Lagrangian density is assumed to be
invariant by the action , i.e., $${\cal L}(g_0 d\s)={\cal L}(d\s)
\quad \quad  \forall g_0\in G,$$  For  jets $d (g_0 s) = g_0 ds$
and, thus: \be \label{gglobal} {\cal L}(d(g_0 \s)) = {\cal
L}(d\s). \ee i.e., $\cal L$ {\em is invariant under a ``global
gauge transformations on $U$''}. Nevertheless, if we consider a
particle field  related by a gauge transformation
$\bar\psi=g_U\psi$, with non-constant $g_U: U\rightarrow G$, then
the action does not commute with the operation of taking the
induced jets, that is: \be \label{gsect} d(g_U \s) \neq g_U d\s .
\ee Thus, if ${\cal L}$ does not depend trivially on the $\theta$
part of the jet,

 \be \label{glocal} {\cal L}(d\bar \psi) \neq {\cal
L}(d\s), \ee  i.e., {\em in general, ${\cal L}$ is not invariant
on the gauge orbit of $\s$.}

\begin{remark}{\em Under our approach, the equality in
(\ref{glocal}) is not necessary by reasons of mathematical
elegance (or by any causality condition). As ${\cal L}(d\s)$ is
assumed to be a quantity physically meaningful, {\em the equality
in (\ref{glocal}) is a requirement of (PGI)$^*$}.}\end{remark}

The known answer to this drawback is that  one is forced to
reconsider the definition of ${\cal L}(d\s)$, and regard the
primary vector fields as a part of the set of all the physical
fields, introducing some new physical fields, to restore
(PGI)$^*$. Remarkably, such fields determine a connection in the
principle bundle $P(M,G)$ (and, thus, in $E(M,V)$). This allows to
define a {\em covariant derivative} $D\s$ which involves the
derivatives of
 $\s$ but is naturally $G$-invariant. That is, the connection
yields a way to generate 1-jets from $\s$ which is free of the
problem (\ref{gsect}) and, thus:
 \be \label{glocals} {\cal L}(D\bar \s) (=  {\cal L}(D(g_U \s)) = {\cal
L}(g_U D\s)) = {\cal L}(D\s), \ee in agreement with (PGI)$^*$.

We can reformulate this physical input as follows: \ben \item[(1)]
The operation of taking jets from a particle field yields
different measures by different standard observers. Thus,
according to (PGI)$^*$, such jets cannot generate  by themselves
physically meaningful quantities  (as the Lagrangian density
function ${\cal L}(d\s)$).\item[(2)] The underlying reason is the
following. If there is no a connection in $P(M,G)$, a standard
observer $\sigma: U\rightarrow P$ does not have any way to compare
the base he chooses at some point $\sigma (p)\in P_p$ with the
bases he chooses at neighboring points. The only known
mathematical element which allows such a comparison is a
connection. \item[(3)] Nevertheless, in general, a particle fibre
bundle has no any canonical connection. So, the introduction of a
connection in $P(M,G)$ is interpreted as the existence of a new
physical field.

In fact, the introduced connection or {\em gauge field}, must be
combined with the differentiation of the section representing the
particle field, in order to yield a well defined operation for
standard observers (covariant derivative). So, the gauge fields
have a interpretation as {\em fields of interactions} with
particle fields. This interpretation (which looks rather ``baggage
dependent''),  may have further consequences for the
theory\footnote{\label{f11} Say, the Lagrangian density not only
is defined by using covariant derivatives, but also must include a
new term which represents the free Lagrangian density for the
gauge fields. (As these fields can be also measured by using
standard observers, the new Lagrangian term must depend on the
curvature of the connection --Utiyama's theorem-- in order to
preserve (PGI)$^*$).}. \een In their own right, these items are
compelling  for the general existence of gauge fields.
Nevertheless, one can wonder at what extent they are unavoidable,
and additional arguments can be provided for each item:

1.- The bad transformation of measures when taking jets, relies on
the computation of the left hand side of (\ref{gsect}), which
involves derivatives of $g_U$. Nevertheless, even under
(\ref{gsect}) some Lagrangian densities could be defined in such a
way that the equality in (\ref{glocal}) holds. However, such
counterexamples are in some sense degenerate and
non-generic\footnote{As a simple example let $M$ be
Lorentz-Minkowski spacetime, $V=\R^2$, $E$ diffeomorphic to
$M\times V$ and the principle bundle $P(M,G)$ be obtained as a
reduction of the bundle of the bases $BE(M,G_2)$ with structural
group: $ G=\{\left(\begin{array}{ll} \lambda  & 0\\0 & 1
\end{array}\right): \lambda>0\}.$ Taking a global section
$\sigma: M\rightarrow P$, we can identify $E \equiv M\times \R^2$,
$P \equiv M\times G$. The gauge orbit of a particle field $\s =
(\s^1,\s^2)$ is the set of all the sections $x\rightarrow
(\lambda_M(x)\s^1(x),\s^2(x)), x\in M$ constructed for any
positive function $\lambda_M$ on $M$. Notice that the directions
of the components $\psi^1, \psi^2$ are gauge independent. Then, a
Lagrangian density which does not depend on $\s^1$  will be gauge
invariant, even if it depends on the derivatives of $\s^2$.}. And
 the usage  of a {\em potentially big} set of scalar quantities
involving derivatives of $\s$, as arbitrary Lagrangian densities,
seems a basic requirement.
 So, item (1) is supported.


2.- The only differential operators defined in arbitrary manifolds
with no additional structure deals with the differential or the
Lie bracket. The latter has the same problems of gauge invariance
than the former (as well as other problems: the Lie bracket must
be applied to pairs of sections, it is defined only for the
tangent bundle and associate tensor bundles, etc.) So, in order to
ensure gauge invariance, an  additional structure must be defined.
Connections in principle fiber bundles (concretely, in reductions
of a fiber bundle of bases) are designed specifically to compare
bases at different points, supporting item (2).

3.- A well-known exception to the inexistence of a unique
canonical connection is the fiber bundle of orthonormal bases for
a semi-Riemannian metric. In fact,  one has a controlled set of
possible connections, and only one torsion free. This
(Levi-Civita) connection is the canonical choice. Then, one can
interpret the Levi-Civita connection as: (i) a gauge field, which
appears in a rather rigid way, or (ii) a pure consequence of a
different structure, the metric. In principle, the difference
between these two ``baggage'' interpretations might yield
consequences for quantities as the Lagrangian term associated to
the connection (footnote \ref{f11}). At any case, Levi-Civita
connection affects only to spacetime (to be discussed in
Subsection \ref{s3.2.3}). If this case (and eventually other
exceptions of principal fiber bundles with a canonical connection)
is excluded, the interpretation (i) of the connection as a field
of interactions appears clearly, in agreement with item (3).

\smallskip

Summing up, with the above precisions:

\begin{quote}
 Assuming  (PGI)$^*$, if $G$ is not a discrete group and enough physical quantities depends on the derivatives of a particle field $\s: M \rightarrow E$, then a connection on $P(M,G)$ must be
included in the geometry of $P$; eventually, such a connection
will be  regarded as a physical field in its own right.
\end{quote}

\subsection{Spacetime theories as Yang-Mills theories}
\label{s3.2.3} Postulates on spacetimes were introduced under a
 viewpoint somewhat  different to gauge theory. But, as
stressed in Section \ref{s3.1}, consensus hypotheses (H1), (H2),
(H3), (H4)$^*$, can be regarded as extensions of (P1), (P2), (P3),
(P4)$^*$. In particular, the principle of gauge invariance
(PGI)$^*$ is also a consensus hypotheses for spacetimes. We
revisit  the conclusions for this case.

(P1), (P2), (P3) were introduced to measure spacetime $M$ itself,
not any field on it. Nevertheless, the tangent vector bundle is
naturally associated to $M$ and, finally, standard observers led
to a submanifold $P$ of the principle bundle of  the bases
$BTM(M,G_4)$. Connections in $BTM(M,G_4)$ can be expressed easily
as a (Koszul) differential operator $\nabla$  on $TM$, as we will
do here.

Under (P4)$^*$ ($k\equiv -c^2$), the submanifold $P$ is the
principle bundle of frames $F(M,O^k(4,\R))$ (orthonormal bases up
to the normalization of the first vector to
 $c$), for a Lorentzian metric $g$. Without loss of generality, we can take
 the fiber bundle of orthonormal frames $F(M,O_1(4, \R)), O_1(4, \R)=O^{k=-1}(4, \R)$.

Let us review briefly this well-known case. The Levi-Civita
connection $\nabla^g$ is automatically selected in $F(M,O_1(4,
\R))$, and any other Koszul connection $\nabla$ on $M$ can be
written as $\nabla_XY = \nabla^g_XY + T_S(X,Y) + T_A(X,Y)$ where
$T_S, T_A$ are vector valued bilinear maps, $T_S$ symmetric and
$T_A$ skew-symmetric. Notice that $T_S, T_A$ are tensor fields
canonically associated to $\nabla$ through $g$.

Accepting that some physical fields can be described by means of a
section on $TM$ or its associated tensor bundles, (PGI)$^*$
applies, and implies the existence of a connection in $F(M,O_1(4,
\R))$. Necessarily this connection is type $\nabla= \nabla^g +
T_A$, where $T_A$ is (up to a factor) the torsion of $\nabla^g$.
Then, it is natural to consider $\nabla^g$ as the gauge field, and
$T_A$ as a new possible particle field (some authors have tried to
use the torsion to describe different physical effects, see for
example \cite{torsion}). An arbitrary connection on $M$, which
does not come a priori from a connection on $F(M,O_1(4, \R))$
(say, as a priori in Palatini method for Einstein equations) would
lie out of the scope of (PGI)$^*$, but would be also equivalent to
consider a particle field $T_S$, in addition to $\nabla^g, T_A$.

Now, assume only (H4)$^*$ instead of (P4)$^*$. Then $k$ is also
constant but non-negative values are permitted. In particular,
when $k=\omega$, the associated Leibnizian structure $\Omega, h$
(the former a non-vanishing one form and the latter a Riemannian
metric on its kernel) is equivalent to a reduction $$GM(M,
O^{\omega}(4,\R)) \subset BTM(M,G_4),$$ where the structural group
$O^{\omega}(4,\R)$ is, {\em the (matrix) Galilean group}.

 (PGI)$^*$ implies the existence of a connection in
the fiber bundle  $GM(M,O^{\omega}(4,\R))$, or {\em Galilean
connection}. Such a connection $\nabla$ parallelizes the
Leibnizian structure ($\nabla \Omega =0, \nabla h = 0$). Galilean
connections always exist, and, in fact, they have as many degrees
of freedom as the connections (with and without torsion) which
parallelize a semi-Riemannian metric. Nevertheless, symmetric
Galilean connections  exist if and only if $\Omega$ is closed
 (for a
complete study of all these questions, see \cite{BS}).

Galilean connections can be reconstructed by means of a
Koszul-type formula from a ``gravitational vector field'' and a
``vorticity field'' (plus the torsion, if non-symmmetric ones are
also considered), \cite[formula (13)]{BS}. But  none of these
tensor fields are selected by our postulates and, thus, neither a
Galilean connection.

So, classical Galilei-Newton theory can be regarded as a (proper)
gauge theory, where the connection is not univocally determined a
priori, but appears from the necessity of gauge
invariance\footnote{In fact, famous Leibniz's objection to
Newton's inertial observers, can be seen as a way to claim for
gauge invariance, and Newton's answer on the spinning water-bucket
might be regarded as the way to postulate the existence of a
connection ---a gauge field for classical space and time.}. In
this sense, the interpretation of the Galilean connection as a
physical gauge field becomes more clear than in the case of
General Relativity.

The cases $k=0, k\in(0,\infty)$ are analogous to the previous ones
--even though they do not have analogs in classical theories.

\section{A further possibility}
\label{s4.1} As we showed in Section \ref{s2}, if one drops
(P4)$^*$ among consensus hypotheses, new possibilities appear,
some of them with interesting interpretations, as the possible
variations of the speed of light. This also suggests the
possibility to remove (H4)$^*$ in general, and explore the
physical and mathematical consequences.

\subsection{Framework for pointwise structural groups} \label{s4.1a}
Recall that,  under (H1), (H2), (H3),  one has the vector fiber
bundle $E(M,V)$, the principle fiber bundle of all the bases
$BE(M,G_m)$ and the subbundle $P$ of $BE(M,G_m)$ containing the
standard bases. As justified in (H2),  some (closed) Lie group
$G_p$ depending on $p$ must act freely and transitively on each
fiber of $P_p$. For simplicity, all the groups will be considered
as connected in what follows.

This space will be denoted $P(M,G_{\star})$. Nevertheless, its
structure is not totally well defined yet, as a precise notion of
smoothability (required by (H3)) must be provided. Clearly, this
notion must include these two items: \ben \item[(i)] $P$ is a
 smooth submanifold embedded in $BE(M,G_m)$ as a closed  subset.

\item[(ii)] Both, $G_p$ an its action on $P_p$ varies smoothly
with $p$. \een

\noindent The first item is  defined accurately, but the second is
not, and will be analyzed in the next subsection. Now, we
emphasize that (i) is not sufficient. 
In fact,  any sensible definition of (ii) would imply that
dim$G_p$ is independent of $p$, but this is not implied by (i):

\bexam {\em Put $M=(-\infty,1), V=\R^2$ and $E=M\times V$, and
construct $P$ as follows. For any $v\in V\backslash\{0\}$, let
$B_v$ the (ordered) base $B_v=(v,w)$ univocally determined by: $
v\cdot w = 0, \parallel v \parallel =  \parallel w
\parallel,$ and $B_v$ is positively oriented
(the usual scalar product $\cdot$, norm $\parallel \cdot
\parallel$ and orientation are considered on $V=\R^2$). Fix $v_0
\in V$ with $
\parallel v_0 \parallel >1$. For each $p\in M$ put:
$$
P_p= \left\{ \begin{array}{lr} \{B_v: \parallel v-v_0 \parallel^2
=
p\} & 0\leq p <1 \\
\{B_v: \parallel v \parallel^2 = 1/|p| & -\infty <p <0 \}
\end{array}\right.
$$
Notice that $P=\cup_{p\in M} P_p$ is a closed embedded submanifold
of dimension 2 in $BE(M,G_2)$. 
Clearly,
$G_p$ is 1-dimensional (in fact, a circle) at any $p\neq 0$, but
it is the (0-dimensional) trivial group at $p=0$. Notice also that
$P$ does not admit any (continuous) local section in a
neighborhood of $p=0$.}\eexam

\subsection{Problem on smoothness and  structure of
$P(M,G_{\star})$}\label{s4.1b}

In order to ensure the smoothness according to (ii) above, one
must ensure that dim$G_p$ is constant, but this condition seems
too weak by itself. Notice that $\{G_p: p\in M\}$ is just a set of
subgroups of $G_m$, with no further structure. A possibility would
be to add an assumption such as ``all the groups $G_p$ are
diffeomorphic''. In fact, this happened in the case of spacetimes
when $k\in (-\infty,0)$,
and this was enough to model variations of speed of light.
Nevertheless, there are are reasons to avoid such an assumption:
(a) it is not justified by first principles, (b)  technical
problems would not be solved in general,  as the diffeomorphisms
between the groups $G_p$ may be non-canonical, and (c) it is a
relatively strong hypothesis and may forbid
 some interesting possibilities.

 In fact, objection (c) would happen if that assumption is imposed to spacetimes.
Notice that the groups $O^{k(p)}(4,\R)$ are all conjugate (and
thus, diffeomorphic) if either $k(p)\in (-\infty,0)$ or $k(p)\in
(0,\infty)$. But the first case corresponds to the Lorentz group,
and the second one to the orthonormal group, which are
topologically very different. Nevertheless, this change of
topology may be interesting, and can occur in a smooth way:

\bexam {\em Let $M=\R^4$, and then $TM\equiv  M\times \R^4$,
$BTM\equiv M\times G_4$. Let $k:\R^4\rightarrow \R$ be any smooth
function with non-constant sign. Let ${\cal M}_4(\R)$ be the set
of square matrixes 4$\times$4 and consider the map $F: M\times G_4
(\equiv BTM) \rightarrow M \times {\cal M}_4(\R) \times \R$
defined as:
$$
F(p,A)=(p, A^t\,I_4^{[k(p)]}\,A , \hbox{det}A).
$$
Recall that $N=\{(p,I_4^{[k(p)]},1): p\in M\} \subset  M \times
{\cal M}_4(\R) \times \R$ is a closed embedded submanifold of the
codomain. Now, choose $P=F^{-1}(N)$.  $P$ has a structure of fiber
bundle $P(M,G_{\star})$ with pointwise fiber $G_p =
O^{k(p)}(4,\R)$ for all $p$. As F has constant rank on $P$, $P$
becomes a smooth submanifold of $BTM$, even at the changes of sign
of $k(p)$ (where the topology of $G_p$ changes).}
 \eexam One can explore some alternatives
for the meaning of smoothability hypothesis (ii).
At any case, from  any reasonable definition of (ii) one would
have: \bit \item[(ii)$_1$] Constancy of the dimension: the
dimension of $G_p$ is independent of $p$.

\item[(ii)$_2$]  Existence of local sections ({\em or standard
observers}): let $\pi: P\rightarrow M$ is the natural projection,
for any $p\in M$ there exists a neighborhood $U\subset M$ and a
map $\sigma: U\rightarrow P$,  such that, $\pi \circ\sigma$ is the
identity of $U$.

\item[(ii)$_3$] Compatibility of sections and actions: given two
sections on $\sigma_1, \sigma_2: U\subset M \rightarrow P$, define
$g_U(p) \in G_p \subset G_m$ by means of $\sigma_2(p) =
\sigma_1(p) g_U(p)$, for all $p\in U$; then both, the map
$g_U:U\rightarrow G_m$ and
$$
\pi^{-1}(U)(\subset P) \rightarrow \pi^{-1}(U), \quad u_p(\in P_p)
\rightarrowtail u_p g_U(p)
$$
 are smooth.
 \eit

\noindent These items can be imposed as a (provisional)
definition. Thus, summing up,
\begin{quote}
For any field theory, (H1), (H2), (H3) imply the  fiber bundle
space $P(M,G_{\star})$.

On each fiber $P_p$, of $P$, a Lie group $G_p$ acts freely and
transitively, and such actions satisfies the requirements of
smoothability (i), (ii) above, the latter implying (ii)$_1$
(ii)$_2$ (ii)$_3$.

In particular, $P(M,G_{\star})$ admits {\em standard observers}.
\end{quote}

\subsection{Gauge invariance} \label{s4.1c}
Notice that, fixing a standard observer $\sigma_0: U\rightarrow P$
one has associate coordinates as in (\ref{ecoor}) and an action of
each $G_p$ on $E_p$ as in (\ref{eact}). Thus, the notions of gauge
transformation and gauge orbit  makes sense and, reasoning as in
in Subsection \ref{ss2.3.1}, one arrives at:

 \ben \item[(PGI)] (Generalized Principle of Gauge Invariance). Under hypotheses (H1), (H2), (H3), all the particle fields in the same orbit Orb($\s$) are physically identical.
 \een
Now, reasons as those in Subsection \ref{s3.2.2} show the
necessity of a geometrical object to compare different bases by a
standard observer. That is,  we must extend the notion of  {\em
connection} to $P(M,G_{\star})$.


Recall that there are different well-known ways to define a
connection in a principe bundle. The definition as a distribution
admits an obvious extension for $P(M,G_{\star})$. Say, a
connection on $P(M,G_{\star})$ is a distribution $H$ ({\em
horizontal distribution}) on $P$ such that: (i) at each $u_p\in
P_p$ the subspace $H_{u_p} \subset T_{u_p}P$ is complementary to
the {\em vertical} subspace determined by vectors tangent to the
fiber $P_p$, and (ii) the distribution is invariant by the action
of $G_p$ at each $p$ ($R_{g_p*}H_{u_p} = H_{u_pg_p}$).

As far as we know, such fibered spaces and connections have not
been studied systematically, even though the possibility to extend
the formalism of principle fiber bundles is well-known \cite{Mu}.
So, we stop here. We emphasize that the mathematical study of such
connections and the possible associated physical phenomenons
in $P(M,G_{\star})$ appears as  questions worth to be studied.
\section{Conclusions}

Finally,  some of the points along this work are emphasized: \ben
\item Our postulates for both, spacetimes and field theories are
truly ``minimum consensus hypotheses on our way of measuring'':
\bit \item they are not only shared by the standard theories but
also they are  apparently unavoidable for physical theories on a
Universe minimally similar to ours, and \item they are based on
minimal symmetries from the experimental viewpoint, expressed in a
fundamental (reasonably baggage free) sense. \eit


\item Gauge invariance is not  an a priori imposition for
mathematical beauty (i.e. to impose that a global gauge invariance
must be also a pointwise -local- one) but a necessity for the
validity of our way of measuring.

 Gauge invariance is also independent of causal relations in the
spacetime. As suggested in \cite{tH}, if a standard observer $O$
``has chosen'' two standard bases $B_p$ and $B_q$ at two causally
related events $p,q\in M$, and one ``changes'' the chosen basis at
$q$, there is no experimental way at $p$ to distinguish if this
change has been carried out --except when a connection  exists.

\item Connections are geometric elements to ensure gauge
invariance. They are required under minimum hypotheses on the
physical magnitudes in field theories (such as the existence of
enough Lagrangians). In most fiber bundles no canonical connection
exist. So, a connection must be introduced under the (baggage)
interpretation of a interaction physical field.

 \item General Relativity as well as
Galilei-Newton theory can be regarded as  gauge theories, in the
same sense than Yang-Mills field theories. That is,  a connection
is required on a vector fiber bundle (in this case, the tangent
bundle or one of its associated tensor bundles), and this
connection is necessary to preserve  gauge invariance under a
pointwise gauge transformation (for a finite-dimensional gauge
subgroup of $G_4$). In both, Galilei-Newton and General
Relativity, there are the same degrees of freedom for the
connections compatible with the underlying spacetime  structure
--i.e, the underlying Lorentzian metric or  ``Leibnizian
structure''. The difference between them (which might affect the
interpretation of the connection as a physical field) comes from
the fact that, for the case of Galilei-Newton, there is no a
preferred connection obtained by imposing the vanishing of the
torsion.

 \item The
possibility to vary the structural group $G_p\subset G_m$ with
$p\in M$ in the bundle structure $P(M,G_\star)$ (in particular,
the variation of $k(p)$ in spacetimes) is allowed by the theory,
and it seems an interesting possibility, as shown in spacetimes:
\bit \item ``Mild'' variations of $G_p$ (as variations as a
conjugate group, which would be equivalent to have a fixed
subgroup of $G_m$ and a different action on each fiber) may model
effects such as the variation in the speed of propagation of
interactions. \item ``Strong'' variations of $G_p$ (including
changes of topology, which can be carried out even in a smooth
way) may describe more drastic effects such as changes of
signature in a metric  of the fiber. \eit Even more, new problems
appear from the purely mathematical viewpoint (existence of
connections on $P(M,G_*)$, associate operators, etc.)

\item The strong mathematical conclusions obtained for both the
spacetime and the field theories can be interpreted from different
viewpoints as:

\bit \item A positivist simplification of our models of Universe.
\item A set of limit requirements of compatibility for any {\em
Theory of Everything}. \item A bound for the number of possible
parallel Universes minimally similar to ours (even selected by the
antropic principle). \item Additionally, for adherents to the
Mathematical Universe Hypothesis \cite{Te} (or related ones, see
\cite{Ba}), a help to descend from the bird to the frog views.
This was explained in \cite{BSS} for the case of spacetimes, and
is suggested now for gauge invariance: among the mathematical
structures compatible with the fiber bundle structure $P(M,G_*)$,
only those with a gauge symmetry can contain observers similar to
ours. \eit \een


\section{Appendix: a sketch of the mathematical development for spacetime postulates}

The mathematical translation of the second postulate (P2) in
Section \ref{s2} is the following. Let
 $ B_p=(\partial_t|_p , \partial_{x^1}|_p, \partial_{x^2}|_p,
$ $\partial_{x^3}|_p),$ $
 \tilde B_p=(\partial_{\tilde t}|_p , \partial_{\tilde{x}^1}|_p, \partial_{\tilde{x}^2}|_p,
\partial_{\tilde{x}^3}|_p)$
be the bases of the tangent space obtained by two standard
observers around  $p$, with transition matrix:
$$ A= \left(
\begin{array}{c|c}
\partial_{\,t}\,{\tilde t}|_p & \partial_{\,x^j}{\tilde t}|_p \\ \hline
\partial_{\,t}{\tilde x}^i|_p   & \partial_{\,x^j}{\tilde x}^i|_p
\end{array}
\right).
$$
Rewriting $A$, equation  (\ref{ep2}) is:
\begin{equation} \label{ecp2}
 A = \left(
\begin{array}{c|c}
 {\mathbf a_{00}} &  a_h \\ \hline
 a^t_v   &  {\mathbf{\hat A}}
\end{array}
\right) \quad \quad \Longrightarrow \quad \quad A^{-1} =\left(
\begin{array}{c|c}
 {\mathbf a_{00}} & {\tilde a}_h \\ \hline
{\tilde a}^t_v  &  {\mathbf{\hat A}^t}
\end{array}
\right),
\end{equation}
where  $\mathbf{a_{00}} \in \R$, ${\mathbf{\hat A}}$ is a
submatrix $3\times 3$ with transpose ${\mathbf{\hat A}^t}$, and
$a_h, a_v, \tilde a_h , \tilde a_v$ are four three-uples of real
numbers ---the Postulate of Time Orientation would yield
$\mathbf{a_{00}}>0.$

The matrixes satisfying (\ref{ecp2}) can be computed directly from
the elementary algorithm to calculate the inverse matrix.
 In order to describe the results, define $O^{(k)}(4,\R)$,  $k\in
\R, k\neq 0$, as the group of the real matrixes
 $4\times 4$ which preserve the matrix
$$
I_4^{[k]}= \left(
 \begin{array}{cccc}
k  & 0 & 0 & 0 \\
0   & 1 & 0 & 0 \\
0   & 0 & 1 & 0 \\
0   & 0 & 0 & 1
\end{array}
\right)
$$
by  congruence, that is:
\begin{equation} \label{ok}
A^t\,I_4^{[k]}\,A = I_4^{[k]}
\end{equation}
or, equally, by taking inverses:
\begin{equation} \label{okprima}
(A^t)^{-1} \,I_4^{[1/k]}\,A^{-1} = I_4^{[1/k]}.
\end{equation}
Definition (\ref{ok}) (resp. (\ref{okprima})) is extended
naturally to the case  $k=0$ (resp $k=\omega$) just by assuming
additionally  det$^2A= 1$. Then, it is not difficult to check the
existence of some $k(p)$ as described in the point (A) of
Subsection \ref{s22}. Moreover, it is easy to identify the four
geometrical structures in $T_pM$ preserved by $O^{(k(p))}(4,\R)$.
These structures are: (1) a Lorentzian scalar product if $k(p)\in
(-\infty, 0)$, (2) a non-zero 1-form $\Omega_p$ and a Euclidean
(positive definite) scalar product $h_p$ in the kernel of
$\Omega_p$ if $k(p)=\omega$, (3) a non-zero vector $Z_p$ and a
(positive definite) scalar product $h^*_p$ in the kernel of $Z_p$
in dual space $T_pM^*$ if $k(p)\in 0$, and (4) an Euclidean scalar
product if $k(p)\in (0, \infty)$. These structures, varying
smoothly with $p$, yield the four structures 1--4 described in the
point (B) of Subsection \ref{s22}.

\section*{Acknowledgments}
The author acknowledges warmly the discussions and comments by
Prof. F. Soler Gil (U. Bremen) and A.N Bernal (U. Granada).

This research has been partially supported by Spanish MEC-FEDER
Grant MTM2007-60731 and Regional J. Andaluc\'{\i}a Grant
P06-FQM-01951.

   

{\small

\begin{thebibliography}{99}

\bibitem{torsion} H.I. Arcos, J.G. Pereira:
Torsion and the gravitational interaction {\em Classical Quantum
Gravity} {\bf 21} (2004), no. 22, 5193--5202.

\bibitem{Ba} J.D. Barrow: {\em Pi in the Sky: Counting, Thinking, and Being}, Oxford Univ. Press, N.Y. (1992).

\bibitem{Ba2} J.D. Barrow: {\em New Theories of Everything}, Oxford Univ. Press, N.Y. (2007).

\bibitem{BSfound} A.N. Bernal, M. L\'opez, M. S\'anchez: Fundamental units of length and time, {\em Found.
Phys.} {\bf 32} (2002), no. 1, 77--108.

\bibitem{BS} A.N. Bernal,  M. S\'anchez: Leibnizian, Galilean and Newtonian structures of space-time, {\em J. Math. Phys.} {\bf 44} (2003), no. 3,
1129--1149.


\bibitem{BSS} A.N. Bernal,  M. S\'anchez, FJ. Soler Gil: {\em Physics from scratch. Letter on M.
Tegmark's ``The Mathematical Universe''}, arxiv: 0803.0944 .

\bibitem{Bl} D. Bleecker: {\em Gauge theory and variational principles}, Global Analysis Pure and Applied Series A, 1. Addison-Wesley Publishing Co., Reading, Mass. (1981).

\bibitem{Dr} T. Dray: General relativity and signature change. Advances in differential geometry and general relativity,
103--124, {\em Contemp. Math.}, {\bf 359}, Amer. Math. Soc.,
Providence, RI, 2004.


\bibitem{HH} J.B. Hartle, S.W. Hawking: Wave function of the Universe,  {\it Phys. Rev. D} {\bf 28} (1983) 2960.

\bibitem{Mu} O. Muller: {\em A metric approach to Fr\'echet geometry},
preprint, arXiv:math/0612379.


\bibitem{MSV} M. Mars, J. M. M. Senovilla, R. Vera:
Lorentzian and signature changing branes, {\em Phys. Rev. D}{\bf
76},  (2007) 044029 (22pp).

\bibitem{Na} G.L. Naber: {\em Topology, geometry, and gauge fields. Interactions.}
Applied Mathematical Sciences, {\bf 141}. Springer-Verlag, New
York, (2000).


\bibitem{Po} W. Poor: {\em Differential geometric structures}, McGraw-Hill Book Co., New York, 1981

\bibitem{Te} M. Tegmark: The Mathematical Universe, {\em Found.
Phys.}, {\bf 38} (2008) 101-150.

\bibitem{Te0} M. Tegmark: Is ``the Theory of Everything'' Merely the Ultimate Ensemble Theory?
{\em Annals of Physics}, {\bf 270} (1998) 1-51

\bibitem{tH} G. 't Hoopf: Gauge theories of the forces between elementary particles, {\em Scientific American} {\bf 242}
(1980) 104-138.

\bibitem{We} S. Weinstein: Gravity and Gauge Theory
in {\em Philosophy of Science}, Vol. {\bf 66}, No. 3, Supplement.
Proceedings of the 1998 Biennial Meetings of the Philosophy of
Science Association. Part I: Contributed Papers. (Sep., 1999), pp.
S146-S155.

\bibitem{YM} C. N. Yang,  R. L. Mills:
Conservation of isotopic spin and isotopic gauge invariance. {\em
Physical Rev.} (2) {\bf 96}, (1954). 191--195.




\end{thebibliography}
}
\end{document}